# Non-standard quasiadditive integrals of motion and pressure dependence of phonon populations

F.S. Dzheparov

NRC "Kurchatov Institute" – ITEP, Moscow, Russia

*The existing equilibrium statistical physics is based on application of standard quasiadditive integrals of motion, which include energy, momentum, rotation momentum, and number of particles. It is shown that this list is far from complete and that any quasiadditive dynamic variable can be mapped to corresponding quasiadditive integral of motion. As a result an ensemble with a given external pressure is constructed. It provides the first example of the distribution in which phonon populations depend on pressure differently than in the canonical Gibbs ensemble.*

### 1. Introduction

Modern equilibrium statistical physics is based on Gibbs distributions [1-4]. To justify them large system consisting of large subsystems is usually considered. Then the assumption of quasi-independence of subsystems produces that the logarithm of the statistical operator of the system is quasi-additive with respect to the logarithms of the statistical operators of subsystems $\rho_a$:

$$\ln \rho = \sum_a \ln \rho_a. \tag{1}$$

This equality is correct when surface effects are ignored. It immediately leads to the canonical Gibbs distribution, since it is assumed that all quasi-additive integrals of motion are reduced to the total energy, the total number of particles, and the total values of the translational and rotation momentums.

The informational view on the problem [5,2,6] is based on the quasi-additivity and maximality of the entropy of the equilibrium distribution. It leads to the same conclusions, since it uses the same integrals of motion to formulate conditions

$$\langle I_\mu \rangle = \mathrm{Tr}\, I_\mu \rho = I_\mu^{(ex)}, \tag{2}$$

limiting the variation $\delta\rho$ in the search for the maximum of the entropy

$$S = -\langle \ln \rho \rangle. \tag{3}$$

As a result

$$\rho = \exp\left(-\sum_{\mu=1}^{\nu} A_\mu I_\mu\right), \tag{4}$$

where $\nu$ is the number of conditions (2) that serve as equations for determining parameters $A_\mu$ from a given set of values $\{I_\mu^{(ex)}\}$. The maximum of entropy produces the distribution corresponding to the minimum of information about the system when conditions (2) are fulfilled.

One of consequences of the equivalence of Gibbs ensembles for non-small systems is that for many phenomena, related to the anharmonicity of lattice vibrations, the dependence of the average numbers of phonon populations $n_k = \langle c_k^+ c_k \rangle$ in the crystal on external pressure $P^{(ex)}$ is manifested only through the pressure dependence of the phonon frequencies $\omega_k\left(P^{(ex)}\right)$ [7,8]:

$$n_k = n_k^{(0)}\left(P^{(ex)}\right) = \left[\exp\left(\beta\omega_k\left(P^{(ex)}\right)\right) - 1\right]^{-1}. \tag{5}$$

Here $\beta = 1/T$ is the inverse temperature, and $c_k^+$ with $c_k$ are the phonon creation and annihilation operators for the state with dimensionless number $k$. Relation (5) forms the basis of the so-called quasi-harmonic approximation [7]: oscillations are small and can be considered as harmonic, and the influ-ence of the lattice anharmonism is reduced to the dependence of the phonon frequencies on the volume of the crystal, associated with the pressure. This approximation is widely used, for example, to describe the thermal expansion and thermal conductivity of crystals.

Direct measurement of the dependence $n_k(P^{(ex)})$ was performed in Refs. [9,10]. The comparing of the intensities of Stokes and anti-Stokes components in Raman scattering was applied. This work continued preceding studies of the authors in Raman light scattering on stressed silicon crystal plates [11]. The results [9,10] are described not by the formula (5), but by the relation

$$n_k(P^{(ex)}) = \left\{\exp\left[\beta\left(\omega_k(P^{(ex)}) + \Delta_k(P^{(ex)})\right)\right] - 1\right\}^{-1}, \qquad (6)$$

where $\Delta_k(P^{(ex)}) = \omega_k(P^{(ex)}) - \omega_k(0)$. The measurements [9,10] were performed at $|\Delta_k(P^{(ex)})| \leq \omega_k(P^{(ex)})/15$ that corresponds to the field of applicability of the quasi-harmonic approximation.

The result (6) was obtained by a highly qualified team in one of the leading physical centers. However, it did not attract widespread attention of researchers, apparently because it did not receive a convincing theoretical justification. Existing statistical physics offers nothing but the relation (5).

Note that the actual measurements [9,10] were carried out in somewhat more complex conditions than isotropic stretching or compression, but we will limit ourselves to this case only to highlight the most important conceptually part of the problem.

It will be shown below, that the list of quasi-additive integrals of motion is far from being exhausted by the standard dynamic variables mentioned above. New statistical operator will be constructed on this basis, which ensures the fulfillment of an additional condition imposed on the pressure in the system. It leads for the first time to phonon populations wich are different from result (5) of canonical Gibbs distributions. The consideration will be fulfilled within the quasi-harmonic approximation applicability.

The new statistical operator does not coincide with the known P-T distribution [2-4], which produces

$$\langle B \rangle_{pT} = \int_0^\infty dV \operatorname{Tr} B \exp\left(-\beta(H + P^{ex}V)\right) / \int_0^\infty dV \operatorname{Tr} \exp\left(-\beta(H + P^{ex}V)\right)$$

for any observable $B$. It corresponds to the inclusion into conditions list (2) the additional requirement not for the pressure, but for the volume of the system. The P-T distribution for large systems is equivalent to usual canonical distribution.

### 2. Pressure as a dynamic variable and an ensemble with a given pressure

We associate with the pressure the dynamic variable

$$Q(p,q) = -\frac{1}{d}\frac{\partial}{\partial \lambda} H(p/\lambda, q\lambda)|_{\lambda=1}. \qquad (7)$$

Here $H(p,q)$ is the Hamiltonian of the system, $d$ is its spatial dimension, $p$ and $q$ denote all impulses and coordinates, and the numerical parameter $\lambda$ is assumed to be equal to one after calculating the derivative. This choice is due to the fact that the average pressure in the canonical ensemble is

$$P^{(ex)} = -\frac{\partial F(\beta,V)}{\partial V} = \frac{1}{V}\sum_n \langle n|Q|n\rangle \exp\left(\beta(F - \langle n|H|n\rangle)\right) = \frac{\langle Q \rangle}{V}. \qquad (8)$$

Here $F$ is the free energy, and $|n\rangle$ are the eigenvectors for the Hamiltonian $H$. Similar relation is valid in the large canonical ensemble. The derivation of Eq. (7) can be found in Refs. [2] and [12]. It is based on the calculation of the pressure operator $P = Q/V = -\partial H/\partial V$ in the case when the complete Hamiltonian of the system $H_{tot} = H(p,q) + U_b(q/L)$, where the term $U_b(q/L)$ describes the influence of the boundary, the volume of the system $V \sim L^d$, and $H(p,q)$ does not contain an explicit dependence on $V$ in the coordinate representation.

In a typical three-dimensional dynamical system with the Hamiltonian

$$H(p,q) = \sum_{j=1}^{N} \frac{p_j^2}{2m} + \frac{1}{2}\sum_{i \neq j} \Phi(\mathbf{q}_{ij}), \tag{9}$$

where $N$ is the number of particles in the system, $m$ is the mass of the particles,, and $\Phi(\mathbf{q}_{ij})$ is the energy of interparticle interaction,

$$Q(p,q) = \frac{1}{3}\left[2\sum_{j=1}^{N} \frac{p_j^2}{2m} - \frac{1}{2}\sum_{i \neq j} \mathbf{q}_{ij} \partial \Phi(\mathbf{q}_{ij})/\partial \mathbf{q}_{ij}\right]. \tag{10}$$

The analysis of equations of motion for densities of standard quasi-additive integrals of motion also leads to similar representations for pressure; see for example [13-16].

Note that only diagonal elements $\langle n|Q|n\rangle$ appear in (8).

If the interaction $\Phi(\mathbf{q}_{ij})$ decreases rapidly enough with growth of $\mathbf{q}_{ij}$, then the operator $Q(p,q)$ is quasi-additive, as the Hamiltonian (9) is. Accordingly, its diagonal part

$$Q_D = \sum_n |n\rangle\langle n|Q|n\rangle\langle n| \tag{11}$$

is also quasi-additive, and, in contrast to $Q(p,q)$, it is the integral of motion.

Indeed, let the system consists of two parts, and, in disregard of the interaction at the boundary,

$$H(p,q) = H_1(p_1,q_1) + H_2(p_2,q_2), \quad Q(p,q) = Q_1(p_1,q_1) + Q_2(p_2,q_2).$$

Here $(p_a,q_a)$ is a set of kinetic momentums and coordinates of the $a$-st subsystem. Accordingly $|n\rangle = |n_1\rangle|n_2\rangle$, where $|n_a\rangle$ is the eigenvector of the Hamiltonian $H_a(p_a,q_a)$. Therefore

$$Q_D = \sum_n |n\rangle\langle n|Q|n\rangle\langle n| =$$
$$= \sum_{n_1 n_2} |n_1\rangle|n_2\rangle(\langle n_1|Q_1|n_1\rangle + \langle n_2|Q_2|n_2\rangle)\langle n_2|\langle n_1| =$$
$$= \sum_{n_1} |n_1\rangle\langle n_1|Q_1|n_1\rangle\langle n_1| + \sum_{n_2} |n_2\rangle\langle n_2|Q_2|n_2\rangle\langle n_2| = Q_{1D} + Q_{2D},$$

that proves the quasi-additivity.

As a result, the statistical operator

$$\rho = \exp\left[\beta\left(G - H - \tau Q_D\right)\right] \tag{12}$$

is the integral of motion and represents the equilibrium distribution that satisfies the requirement of quasi-additivity (1) and additional conditions

$$\langle H\rangle = E, \quad \langle Q/V\rangle = P^{(ex)}, \quad \langle 1\rangle = 1, \tag{13}$$

which define $\beta$, $\tau$ and $G(\beta,\tau)$. It is obvious that $\langle Q_D\rangle = \langle Q\rangle$.

The relation (12) is also valid in the classical theory for

$$Q_D(p,q) = \lim_{\vartheta \to \infty} \int_{-\vartheta}^{\vartheta} \frac{dt}{2\vartheta} Q(p_c(t), q_c(t)),$$

where $p_c(t)$ and $q_c(t)$ represent the classical trajectory with the initial condition $p_c(t=0) = p$, $q_c(t=0) = q$.

### 3. Equilibrium statistical operators with nonstandard quasi-additive integrals of motion

The above method of constructing a quasi-additive integral of motion $Q_D$ is completely general, and it is not tied to the relation (10) which defines $Q$. It allows to construct a quasi-additive integral of motion $Q_D^{(\mu)}$ for any quasi-additive dynamic variable $Q^{(\mu)}(p,q)$ as the diagonal part of $Q^{(\mu)}(p,q)$. New quasi-additive integrals of motion, obtained in this way, can be further included in the list of additional conditions (2) for constructing new statistical operators according to the rule (4) and for maximizing the entropy $S = -\langle \ln \rho \rangle$.

### 4. Phonon filling numbers in the presence of constant external pressure

Consider a crystal consisting of atoms of the same type. We write, according to [1], the corresponding lattice Hamiltonian in the approximation that is quadratic in the displacements $\mathbf{u}_{\mathbf{n}s}$ of the atoms from their average positions $\mathbf{r}_{\mathbf{n}s}$ in the cell with the number $\mathbf{n}$:

$$H(p,u,V) = \sum_{\mathbf{n}s} \frac{p_{\mathbf{n}s}^2}{2m} + \frac{1}{2} \sum_{\mathbf{nn}'ss'} \Lambda_{ss'}^{\alpha\beta}(\mathbf{n}-\mathbf{n}',V) u_{\mathbf{n}s}^{\alpha} u_{\mathbf{n}'s'}^{\beta} + U_0(V). \quad (14)$$

The index $s$ lists the atoms in the unit cell, and $\alpha$ and $\beta$ numbers the Cartesian components. In (14), in contrast to (9) and due to the application of the harmonic approximation, the energy of the mean positions $U_0(V)$ and the coefficients $\Lambda_{ss'}^{\alpha\beta}(\mathbf{n}-\mathbf{n}',V)$ depend explicitly on the volume $V$.

With this in mind, we get,

$$Q(p,u,V) = -\left(\frac{1}{d}\frac{\partial}{\partial \lambda} + V\frac{\partial}{\partial V}\right) H(p/\lambda, u\lambda, V)|_{\lambda=1} =$$

$$= \frac{2}{d}\left[\sum_{\mathbf{n}} \frac{p_{\mathbf{n}}^2}{2m} - \frac{1}{2}\sum_{\mathbf{nn}'ss'} \Lambda_{ss'}^{\alpha\beta}(\mathbf{n}-\mathbf{n}',V) u_{\mathbf{n}s}^{\alpha} u_{\mathbf{n}'s'}^{\beta}\right] - \quad (15)$$

$$- \frac{1}{2}\sum_{\mathbf{nn}'ss'} \frac{\partial \Lambda_{ss'}^{\alpha\beta}(\mathbf{n}-\mathbf{n}',V)}{\partial \ln V} u_{\mathbf{n}s}^{\alpha} u_{\mathbf{n}'s'}^{\beta} - \frac{\partial U_0(V)}{\partial \ln V}.$$

When recording using the phonon creation $c_k^+$ and annihilation $c_k$ operators for the state $k$, the Hamiltonian (14) passes into

$$H = \sum_k \omega_k \left(c_k^+ c_k + 1/2\right) + U_0(V). \quad (16)$$

The operator $Q$ in this representation looks more complex, but the diagonal part is simple:

$$Q_D = -\frac{\partial}{\partial \ln V}\left(U_0(V) + \frac{1}{2}\sum_k \omega_k\right) - \sum_k \frac{\partial \omega_k}{\partial \ln V} c_k^+ c_k. \quad (17)$$

The results indicated in textbooks [7,8] for pressure within the canonical ensemble lead to the same form for $Q_D$.

Now the relation (12) naturally produces

$$n_k = \left\{\exp\left[\beta\left(\omega_k - \tau \partial \omega_k / \partial \ln V\right)\right] - 1\right\}^{-1}. \quad (18)$$

Here $V = V_0(P^{(ex)})$. The method for calculating this value is described below.

Calculations with the statistical operator (12), as well as calculations in the P-T ensemble, include integration over the system volume, for example,

$$\exp(-\beta G) = \Omega^{-1} \int_0^\infty dV \, \text{Tr} \exp\left[-\beta(H + \tau Q_D)\right], \quad (19)$$

where a formal parameter $\Omega$ with the volume dimension is introduced so that the distribution (12) is dimensionless. It is convenient to calculate the volume integrals in typical calculations of such quantities as $G$, $\langle H \rangle$, $P = \langle Q/V \rangle$ and $n_k$ after calculating the trace by other degrees of freedom. In this case, as in the P-T-ensemble, the result for large systems is determined by the presence of a sharp maximum in $V$ near a certain value $V_0$, and the relative width of this maximum is small in the parame-ter $N^{-1/2}$, where $N$ is the number of particles in the system.

According to the rules of the steepest descent method, the value $V = V_0$ is determined by the equation

$$\frac{\partial}{\partial V} \ln \mathrm{Tr} \exp\left[-\beta(H + \tau Q_D)\right] = \beta\left(P_1 - \tau \left\langle \frac{\partial Q_D}{\partial V} \right\rangle_1\right) = 0. \tag{20}$$

Here $P_1 = -\langle \partial H / \partial V \rangle_1$, $\langle B \rangle_1 = \mathrm{Tr}\, B \rho_1$ for any operator $B$, and

$$\rho_1 = \exp\left[-\beta(H + \tau Q_D)\right] / \mathrm{Tr} \exp\left[-\beta(H + \tau Q_D)\right]. \tag{21}$$

Taking into account (13) we have $P_1 = P^{(ex)}$. Therefore, if $P^{(ex)} = 0$, then Eq.(20) produces $\tau = 0$, and distribution (21) is reduced to the canonical one with the volume $V = V_T$, where $V_T$ is usual equilibrium volume in the absence of external pressure.

In general, $\tau$ and $V_0$ are defined by a pair of relations

$$P_1(V_0, \tau) = P^{(ex)}, \qquad \tau \langle \partial Q_D / \partial V \rangle_1 = P^{(ex)}. \tag{22}$$

The main experimental data in Refs [9] and [10] are connected with deformations significantly greater than the thermal change in the cell parameter associated with the last term of (17) [7,8]. Therefore, this term can be ignored when calculating $\tau$. As a result

$$\tau = P^{(ex)} / \left\langle \frac{\partial Q_D}{\partial V} \right\rangle_1 \approx P^{(ex)} / \frac{\partial(V_0 P^{ex})}{\partial V_0} \approx (V_0 - V_T) / V_T. \tag{23}$$

It was taken into account here that $|V_0 - V_T| / V_T \ll 1$, and $V_0 - V_T \sim P^{(ex)}$.

Now, using the relative smallness of the change in phonon frequencies, we obtain

$$n_k(P^{(ex)}) = \left\{\exp\left[\beta\left(\omega_k(P^{(ex)}) - \Delta_k(P^{(ex)})\right)\right] - 1\right\}^{-1} = $$
$$= \left\{\exp\left[\beta\left(\omega_k(P^{(ex)} = 0)\right)\right] - 1\right\}^{-1}. \tag{24}$$

### 5. Temperature

The relation of the parameters $\beta$ and $\tau$ introduced in the new distribution to the standard thermodynamic temperature $T_t = 1 / \beta_t$ requires special investigation.

In the case of an ideal gas, the pressure operator is proportional to the Hamiltonian: $P = 2H / (dV)$. In this case, the relation (12) coincides with the usual canonical distribution and $\beta_t = \beta(1 + 2\tau / d)$. We see that we have no new ensemble for the ideal gas.

For phonon systems, the new distribution can not be reduced to the previously known ones. Therefore, we first consider the classical limit when $\max \omega_k \ll T_t$. It is natural to expect that the temperature is determined by the average value of the kinetic energy per oscillating atom: $\langle \sum_{j=1}^{N} p_j^2 / 2m \rangle / N = d T_t / 2$. A simple calculation in this case gives

$$\beta_t = \beta(1 + \tau \gamma_{eff}^c). \tag{25}$$

Here we introduced the effective Grűneisen parameter $\gamma_{eff}^c$ associated with the standard [7,8] modal Grűneisen parameters $\gamma_k = -\partial \ln \omega_k / \partial \ln V$ by the relation

$$(1 + \tau \gamma_{eff}^c)^{-1} = \sum_k (1 + \tau \gamma_k)^{-1} / (dN). \tag{26}$$

In the general situation, it is natural to consider the thermometer as separate macroscopic subsystem with known properties. Let the thermometer be an ideal gas with a given volume $V_t$, Hamilto-nian $H_t$, and temperature $T_t = 1/\beta_t$, which is in equilibrium with the crystal described by the density matrix (21). Consider the evolution of a three-dimensional system when the weak interaction between the thermometer and the crystal is switched on. We assume that initially, at $t = 0$, the density matrix of the system

$$\rho_g = \rho_1(\beta, \tau) \rho_t(\beta_t, V_t). \tag{27}$$

Standard theory of linear reaction [2] indicates that the rate of change in the energy of the thermometer

$$\tfrac{\partial}{\partial t} \langle H_t(t) \rangle = \int_0^\infty dt' \operatorname{Tr}[H_1(t'), H_t][H_1, \rho_g]. \tag{28}$$

Here $H_1(t) = \exp(iH_g t) H_1 \exp(-iH_g t)$, and Hamiltonian $H_g = H + H_t$. The relation (28) is valid at the beginning of evolution, when $|\langle H_t(t) \rangle - \langle H_t(0) \rangle| \ll |\langle H_t(0) \rangle|$, but $t > \tau_c$, where $\tau_c$ is the decay time of the integrand. We rewrite (28) using the eigenvectors $|n_g\rangle = |n\rangle|n_t\rangle$ and eigenvalues $E_{gn} = E_n + E_{tn}$ of the Hamiltonian $H_g$, constructed from the eigenvectors and eigenvalues of the Hamiltonians $H$ and $H_t$:

$$\tfrac{\partial}{\partial t} \langle H_t \rangle = \sum_{n_g m_g} \pi \delta(E_{gn} - E_{gm}) |\langle n_g | H_1 | m_g \rangle|^2 (E_{tm} - E_{tn}) \cdot$$
$$\cdot \{1 - \exp[-(\beta_t - \beta)(E_n - E_m) + \beta\tau(Q_{Dn} - Q_{Dm})]\} \rho_{gn}. \tag{29}$$

Here $\rho_{gn} = \exp(-\beta(E_n + \tau Q_{Dn}) - \beta_t E_{nt}) / \operatorname{Tr} \exp(-\beta(H + \tau Q_D) - \beta_t H_t)$ is the eigenvalue of the density matrix for the eigenvector $|n_g\rangle = |n\rangle|n_t\rangle$.

It is natural to expect that when the thermometer is in equilibrium with the crystal
$$\partial \langle H_t(t) \rangle / \partial t = 0. \tag{30}$$

If $\tau = 0$, then this condition is satisfied for $\beta = \beta_t$ regardless of the explicit form $H_1$, as it should be in the canonical ensemble.

In general the condition (30) is an equation on $\beta = \beta(\beta_t, \tau)$, the solution of which depends on $H_1$. There exists natural approximation for this solution, which does not depend on $H_1$. Put $H = H^{(el)} + H^{(ph)}$ and $Q_D = Q_D^{(el)} + Q_D^{(ph)}$. Here, the phonon parts $H^{(ph)}$ и $Q_D^{(ph)}$ include terms that are quadratic in operators $c_k^+$ and $c_k$, and the elastic parts $H^{(el)}$ and $Q_D^{(el)}$ do not contain operators. It is obvious that the density matrix (21) does not depend on $H^{(el)}$ and $Q_D^{(el)}$, since their contributions to the numerator and denominator in (21) are cancelled. Similarly, we can put

$$\rho_1 = \exp\left[-\beta\left((1 + \tau u)H^{(ph)} + \tau(\Delta Q_D^{(ph)} - u\Delta H^{(ph)})\right)\right] / Z, \tag{31}$$

where $\Delta B = B - \langle B \rangle_1$, and $Z$ is the normalization factor. Choose the parameter $u$ to minimize $\Phi = \langle (\Delta Q_D^{(ph)} - u\Delta H^{(ph)})^2 \rangle_1$. At that

$$u = \langle \Delta Q_D^{(ph)} \Delta H^{(ph)} \rangle_1 / \langle (\Delta H^{(ph)})^2 \rangle_1. \tag{32}$$

Now, in the main order by $\Delta Q_D^{(ph)} - u\Delta H^{(ph)}$, we obtain $\rho_1 = \exp\left[-\beta(1 + \tau u)H^{(ph)}\right] / Z$, and the equation (30) acquires the solution

$$\beta = \beta_t / (1 + u\tau), \tag{33}$$

which does not depend on $H_1$. Calculation by the relations (32) and (33) leads to

$$u = \sum_k \gamma_k \omega_k^2 n_k (n_k + 1) / \sum_k \omega_k^2 n_k (n_k + 1), \qquad (34)$$

Where, as in (18), $n_k = \{\exp[\beta(1+\tau\gamma_k)\omega_k] - 1\}^{-1}$. As a rule $|\gamma_k| \sim 1$ [7]. To identify the effects of the main order in $\tau \ll 1$, it is sufficient to put $\tau = 0$ in (34). In this limit, $u$ coincides with the socalled full (or average) Grűneisen parameter $\gamma$ for which there are many numerical estimates in the literature. Apparently, the value $\gamma \approx 0.5$ given in the Ref. [17] for silicon at room temperature did not undergo significant changes in subsequent works [18,19].

In the classical limit $u = \gamma_{eff}^c (1 + O(\tau))$ that coincides with preceding exact result for classical statistics.

### 6. Conclusion

Our construction of the new ensemble is based on the existence of explicit relation for the new quasi-additive integral of motion in the phonon representation for crystal vibrations. Apparently, it can be easily generalized to any other objects for which the Hamiltonian of free quasiparticles is a good approximation, for example, to magnon systems. In a more general situation, the problem may be as complex as the ergodic problem.

The obtained new dependence of the phonon filling numbers on the external pressure (24) differs significantly from that for the previously known Gibbs ensembles (5), but does not agree with the result (6) of the experiment [9,10], even taking into account the temperature redefinition according to Eq.30 discussed above. This disagreement indicates the need for further experimental and theoretical studies. In particular, it becomes much more important to study the correspondence between the experiment and the ensemble related to it. Note that along with optical and magnetic resonance methods for measuring phonon and magnon populations, it is desirable to use inelastic neutron scattering, since in this case the influence of effects such as a narrow phonon bottleneck is minimized.

Currently, in the absence of ergodic theorems, kinetic equations are usually formulated in such a way that their equilibrium solutions coincide with the results of the equilibrium theory. In fact, this requirement is purely phenomenological, and the example exists [20] of how a consistent theory can be in conflict with it.

The extension of the class of permissible equilibrium states identified in our study, which occurs when taking into account non-traditional quasi-additive integrals of motion, should lead to the modification of not only equilibrium, but also non-equilibrium statistical physics. First of all, this may be valid in the problems of the theory of destruction of materials. They were the focus of the experimental works [9-11] that stimulated our research.

To apply our results in the general theory of kinetic equations, the problem of finding convenient representations for densities of non-standard quasi-additive integrals of motion must be solved first. They will be used, toghether with standard densities, in constructing, for example, a description of the hydrodynamic stage of evolution.

Note that the finding of a new quasi-additive integral of motion in the quasiharmonic approximation does not mean that it will appear in the fully equilibrium density matrix of the system. For the theoretical solution of such a question, it would be necessary to prove the corresponding ergodic theorem. But the new integral is at least important in quasi-equilibrium distributions describing the system at the stage of approaching full equilibrium. Similar situations are well known. For example the evolution and many effects of isolated spin systems are described under the assumption that there is an equilibrium within the Zeeman subsystems and dipole subsystem, each of which has its own temperature, and their Hamiltonians are integrals of motion in disregard of the nonsecular terms of the dipole-dipole interactions (see for example [21]). Nevertheless there is no doubt that the complete equilibrium of isolated spin system corresponds to the canonical distribution with single temperature and complete Ham-iltonian, which is exact integral of motion.

I am grateful to I.V. Volovich, D.V. Lvov, S.V. Stepanov, A.N. Tyulyusov, V.G. Hamdamov and V.E. Shestopal for useful discussions.

Russian version of the work can be found in Ref. [22].